\documentclass[aps,pre,twocolumn,showpacs,superscriptaddress]{revtex4}
\usepackage{psfrag}
\usepackage{amssymb,amsmath,amsthm}
\usepackage[dvips]{graphicx,color,epsfig}
\usepackage{verbatim}

\begin{document}
\title{SYNCHRONIZATION INDUCED BY INTERMITTENT VERSUS PARTIAL DRIVES IN CHAOTIC SYSTEMS}
\author{O. ALVAREZ-LLAMOZA}
 \affiliation{Departamento de F\'{\i}sica, FACYT, Universidad de
        Carabobo, Valencia, Venezuela}
\affiliation{Centro de F\'isica Fundamental, Universidad de Los Andes, M\'erida, Venezuela.}
\author{M. G. COSENZA}
\affiliation{Centro de F\'isica Fundamental, Universidad de Los Andes, M\'erida, Venezuela.}
\date{\today}

\begin{abstract} 
We show that the synchronized states of two systems of identical chaotic maps 
subject to either, a common drive that acts with a probability $p$ in time or to the same drive acting on a fraction $p$ of the maps, are similar. The synchronization behavior of both systems can be inferred by considering the dynamics of a single chaotic map driven with a probability $p$. The synchronized states for these systems are characterized on their common space of parameters. Our results show that the presence of a common external drive for all times is not essential for reaching synchronization in  a system of chaotic oscillators, nor is the simultaneous sharing of the drive by all the elements in the system. Rather, a crucial condition for achieving synchronization is the sharing of some minimal, average information by the elements in the system over long times.
\end{abstract}

\maketitle

Chaos synchronization has attracted much
interest from both scientists and engineers by providing
insights into natural phenomena and motivation for practical
applications in communications and control [Pecora \&  Carroll, 1990;  Boccaletti \textit{et al.}, 2002; Uchida \textit{et al.}, 2005; Argyris \textit{et al.}, 2005; Pikovsky \textit{et al.}, 2002].
This phenomenon is commonly observed in 
unidirectionally coupled systems, where a distinction can be made between 
a drive or forcing subsystem and another driven or response subsystem
that possesses chaotic dynamics [Pikovsky \textit{et al.}, 2002].
Complete synchronization occurs when the state variables 
of the driving and the response subsystems converge to a single trajectory in phase space.
On the other hand, generalized synchronization of chaos arises when a functional relation different from the identity is established between the drive and the response subsystems [Rulkov \textit{et al.}, 1995; Abarbanel \textit{et al.}, 1996; Kapitaniak \textit{et al.}, 1996; Hunt \textit{et al.}, 1997; Parlitz \&  Kokarev, 1999;  Zhou \& Roy, 2007]. 

Periodic, chaotic, or stochastic drives have been shown to induce generalized synchronization in chaotic systems [Maritan \& Banavar, 1994; Pikovsky \textit{et al.}, 2002]. 
The auxiliary system approach [Abarbanel \textit{et al.}, 1996] shows that when a response and a replica
subsystems are driven by the same
signal, then the orbits in the phase spaces of the
response and replica subsystems become identical and they
can evolve on identical attractors, if their initial conditions lie on the same basin of attraction of the driven-response system. By extension, an ensemble of identical chaotic oscillators can also be synchronized by a common drive. The specific functional form of the drive is not essential; the basic mechanism that leads to synchronization is the sharing of the same information by the oscillators for all times. In fact, it is has recently been shown that the source of the common influence being received by the elements in an extended system is irrelevant; it could consist of an external drive, or an autonomous global interaction field [Alvarez-Llamoza \& Cosenza, 2008]. At the local level, each element in the system is subject to a source that eventually induces complete or generalized synchronization between the source and the element. 

In this paper, we explore another mechanism for synchronization of a system of driven chaotic elements. We consider a system of chaotic elements where the external drive acts intermittently on all the elements with a probability $p$.  From the analysis of the dynamics at the local level, we extend the auxiliary system approach to a situation where the drive is applied only to a fraction $p$ of randomly chosen elements in a system. We show that the complete synchronized states in both, the intermittently and the partially driven systems, are equivalent. Our results show that the presence of a common drive for all times is not indispensable for reaching synchronization in an extended system of chaotic oscillators, nor is the simultaneous sharing of the drive by all the elements in the system.
Our work is motivated by the practical aspect of searching for minimal requirements for the emergence of synchronization in dynamical systems.

We search for minimal conditions for the occurrence of synchronization of chaos by using models of coupled maps. Let us consider a system of $N$ uniformly, intermittently driven maps, defined as
\begin{equation}
{\begin{array}{l}
\forall i, \; x^i_{t+1} = \begin{cases}
s(x^i_t,y_t), &    \text{with probability $p$}\\
f(x_t^i),  &   \text{with probability $(1-p)$}
            \end{cases} \\
y_{t+1} =  g(y_t),
\end{array}}
\label{int_drive}
\end{equation}
where $x^i_t$ ($i=1,2,\ldots,N$) gives the state of the $i$th
map at discrete time $t$, $\epsilon$ is the strength of the coupling to the drive $g(y_t)$.
Each map is subject to the same external influence (or lack of it) at any time.
The coupling function is chosen to have the diffusive form
\begin{equation}
\label{cfunc}
  s(x_t^i,y_t)=(1-\epsilon)f(x_t^i)+ \epsilon g(y_t) \, .
\end{equation}
We assume a chaotic driven dynamics given by a singular map belonging to the family $f(x_t)=b+| x_t |^z$ ,
where $|z| < 1$, $b$ is a real parameter and $x_t \in (-\infty, \infty)$. These singular maps
exhibit robust chaos, with no periodic windows in a finite interval of the  parameter $b$ [Alvarez-Llamoza \textit{et al.}, 2008]. Robustness is an important property in applications that require reliable operation under chaos in the sense that the chaotic behavior cannot be destroyed by small perturbations of the system parameters.
In particular, we employ the value $z=-0.5$ for which the corresponding map displays
robust chaotic dynamics in the range $b \in [0.62996,1.88988]$.

\begin{figure}[h]
\centerline{
\includegraphics[scale=0.8]{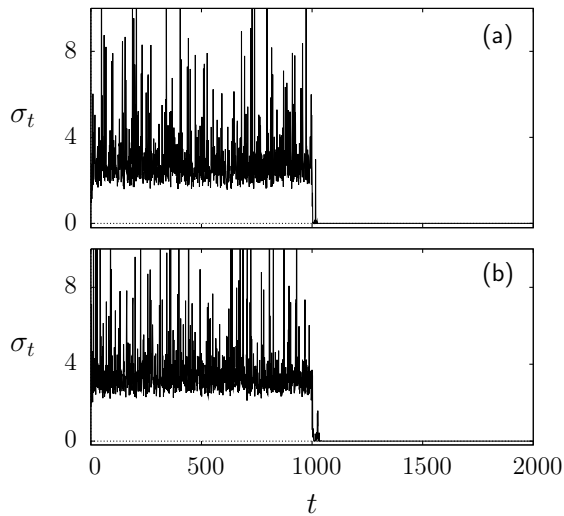}}
\caption{$\sigma_t$ vs. $t$ for the system of driven maps Eq.~(\ref{int_drive}) with  $N=10^4$ and $f(x)=1.2+|x|^{-0.5}$, for different forms of the drive $g(y_t)$. Initial conditions were randomly and uniformly distributed such that $x_0^i \in [0,10]$. The drive $g(y_t)$ is applied with probability $p$ starting at $t=1000$.
(a) $g(y_t)=1.2+|y_t|^{-0.5}$, $\epsilon=0.7$,  $p=0.6$ (complete synchronization). (b) $g(y_t)=1-2y_t^2$, $\epsilon=0.8$,  $p=0.8$ (generalized synchronization).}
\label{fig1}
\end{figure}

A completely synchronized state in the system Eq.~(\ref{int_drive}) is given by $x_t^i=x_t=y_t$, $\forall i$, and it can occur when $g=f$. On the other hand, if $g\neq f$, generalized synchronization, characterized by the condition $x_t^i=x_t \neq y_t$, 
may also arise in this system for $p \leq 1$.
A synchronized state can be characterized by the asymptotic time-average $\langle\sigma\rangle$ (after discarding a number of transients) of the instantaneous standard deviations
$\sigma_t$ of the distribution of map variables $x^i_t$, defined as
\begin{equation}
\sigma_t=\left[ \frac{1}{N} \sum_{i=1}^N \left( x^i_t - \langle x_t \rangle \right)^2 \right]^{1/2},
\end{equation}
where $\langle x_t \rangle$ is the instantaneous mean of the values $x^i_t$, $\forall i$. Stable synchronization corresponds to $\langle \sigma \rangle=0$. Here we use the numerical criterion $\langle \sigma \rangle < 10^{-7}$. Figure~\ref{fig1} shows 
$\sigma_t$  as a function of time for the intermittently driven system  Eq.~(\ref{int_drive}) subject to different chaotic drives $g$. 

\begin{figure}[h]
\centerline{
\includegraphics[scale=0.8]{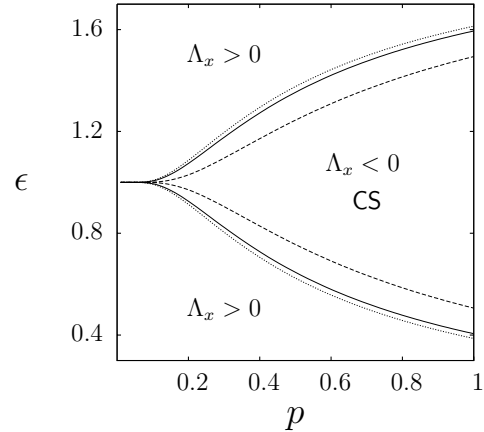}}
\caption{Regions for completely synchronized (CS) states ($\langle \sigma \rangle < 10^{-7}$, after discarding $10^4$ iterates) on the plane $(p,\epsilon)$ for the system of intermittently driven maps, Eq.~(\ref{int_drive}) with $f(x)=1.2+|x|^{-0.5}$, $N=10^4$. Dashed line: $g(y_t)= \overline{x}_1=-0.393713$; solid line: $g(y_t)=1.2+|y_t|^{-0.5}$; dotted line: $g(y_t)=\{ \overline{x}_1=0.204805, ~ \overline{x}_2=-1.00968 \}$.  The boundaries also indicate the onset of stability of those same complete synchronized states in the partially driven system, Eq.~(\ref{part_drive}). The boundaries that separate the stable from the unstable regions are given by the corresponding curve $\Lambda_x = 0$ (Eq.~(\ref{lyapexp})) for the driven single map, Eq.~(\ref{2dmap}).}
\label{fig2}
\end{figure}

For a given value of the coupling strength $\epsilon$, there is a threshold value of the probability $p$ required to reach
either type of synchronization. 
Figure~\ref{fig2}  shows the regions for the complete synchronized states for the system Eq.~(\ref{int_drive}) on the space of parameters $(p,\epsilon)$ for different orbits of a drive $g=f$. In particular, when $g=f$, complete synchronization into an unstable period-$m$ orbit of the
map $f$, defined by $f^{(m)}(\overline{x}_n)=\overline{x}_n$ and satisfying 
$\prod^m_{n=1}
|f'(\overline{x}_n)|>1$, where
$\{\overline{x}_1,\overline{x}_2,\ldots,\overline{x}_m\}$ are the set of
consecutive points on this orbit, can also be achieved in the system Eq.~(\ref{int_drive}), as shown in Fig.~\ref{fig2}.

\begin{figure}[b]
\centerline{
\includegraphics[scale=0.8]{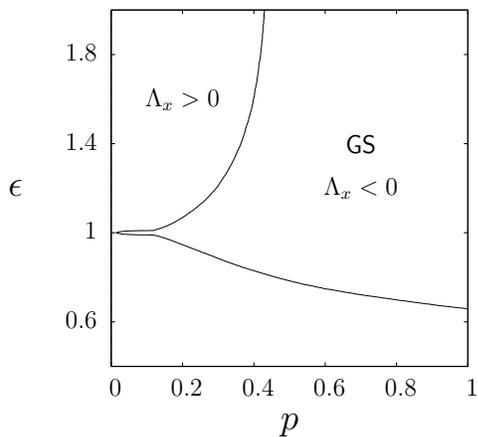}}
\caption{Region for generalized synchronized (GS) states ($\langle \sigma \rangle < 10^{-7}$, after discarding $10^4$ iterates) for the system of maps Eq.~(\ref{int_drive}),
with $f(x)=1.2+|x|^{-0.5}$, $N=10^4$, subject to the intermittent drive  $g(y_t)=1-2 y_t^2$, on the plane $(p,\epsilon)$.  
This same region corresponds to generalized synchronization (satisfying $\Lambda_x<0$) for the single driven map  Eq.~(\ref{2dmap}) with $g(y_t)=1-2 y_t^2$.}
\label{fig3}
\end{figure}

Figure~\ref{fig3} shows the regions for the generalized synchronized states of the system Eq.~(\ref{int_drive})
on the space of parameters $(p,\epsilon)$, with a drive  $g \neq f$. 

Since each map in the intermittently driven system Eq.~(\ref{int_drive}) experiences the same 
external influence (or none) at any time, the properties of this system can be analyzed from the behavior of the individual local dynamics. Thus, we consider a single, intermittently driven map
\begin{equation}
{\begin{array}{l}
x_{t+1} = \begin{cases}
s(x_t,y_t), & \text{with probability $p$}\\
f(x_t),  & \text{with probability $(1-p)$},
          \end{cases} \\
y_{t+1} =  g(y_t),
\end{array}}
\label{2dmap}
\end{equation}
where $x_t$ is the driven or response variable, $y_t$ is the drive and $ s(x_t,y_t)$ has the same functional form as in Eq.~(\ref{cfunc}). The auxiliary system approach [Abarbanel \textit{et al.}, 1996] implies that a driven map can synchronize on identical orbits with another, identically driven map. Thus, the occurrence of stable synchronization in the single map Eq.~(\ref{2dmap}) should lead to synchronization in the extended system of maps Eq.~(\ref{int_drive}), even when the drive acts intermittently in both cases.

The single driven map  Eq.~(\ref{2dmap}) can be regarded as a two-dimensional system. The linear stability condition for synchronization requires the knowledge of the Lyapunov exponents for such system. 
These are defined as $\Lambda_x = \lim_{T\rightarrow \infty} \ln L_x$ and $\Lambda_y = \lim_{T\rightarrow \infty} \ln L_y$, where $L_x$ and $L_y$ are the magnitude of the eigenvalues of $[\prod^{T-1}_{t=0} \mathbf{J}(x_t,y_t)] ^{1/T}$, and $\mathbf{J}(x_t,y_t)$ is the Jacobian matrix for the system  Eq.(\ref{2dmap}), calculated along an orbit. A given orbit $\{x_t,y_t\}$ from $t=0$ to $t=T-1$ can be separated in two subsets,
according to the source of the $x_t$ variable, either coupled or uncoupled to the drive, that we respectively denote as 
$A=\{ \{x_t,y_t\}:  x_t=s(x_{t-1},y_{t-1}) \}$ possessing $pT$ elements, and $B=\{ \{x_t,y_t\}: x_t=f(x_{t-1}) \}$ having $(1-p)T$ elements. We get
\begin{equation}
\left( \prod^{T-1}_{t=0} \mathbf{J}\right)^{1/T} = \left( \begin{array}{cc}
\displaystyle{ \prod_{t: \, x_t \in A} s_x \prod_{t: \, x_t \in B} f'(x_t)} & K\\
0 & \displaystyle{\prod^{T-1}_{t=0} g'(y_t)}
\end{array} \right)^{1/T},
\label{jacobians2}
\end{equation}
where $s_x=\frac{\partial s}{\partial x}=(1-\epsilon)f'(x)$, and $K$ is a polynomial whose terms contain products of $s_x$, $\epsilon$, and $g'(y_t)$ to be evaluated along time.
Then $L_x=[ \prod_{x_t \in A} s_x \prod_{x_t \in B} f'(x_t)]^{1/T}$
and $L_y=[ \prod^{T-1}_{t=0} g'(y_t) ]^{1/T}$. Thus we get 
\begin{equation}
\Lambda_x= p \ln|1-\epsilon| + \lim_{T \rightarrow \infty} \dfrac{1}{T}  \left[ \ln \prod_{x_t \in A} \vert f'(x_t)\vert + \ln  \prod_{x_t \in B} \vert f'(x_t)\vert \right]              
\end{equation}
\begin{equation}
 \Lambda_y =\lim_{T \rightarrow \infty} \dfrac{1}{T} \sum^{T-1}_{t=0} \ln | g'(y_t) | = \lambda_g \, ,
\end{equation}
where $\lambda_g$ is the Lyapunov exponent of the map $g(y_t)$. Synchronization  occurs if the Lyapunov exponent corresponding to the driven map is negative [Rulkov \textit{et al.}, 1995]; i.e., $\Lambda_x<0$. 
For a given set of parameter values, there is a definite value of the probability $p$ at which the exponent $\Lambda_x$ changes its sign, from positive to negative, signaling the onset of synchronization in the dynamics of the two-dimensional system Eq.~(\ref{2dmap}). 

\begin{figure}[h]
\includegraphics[scale=0.8]{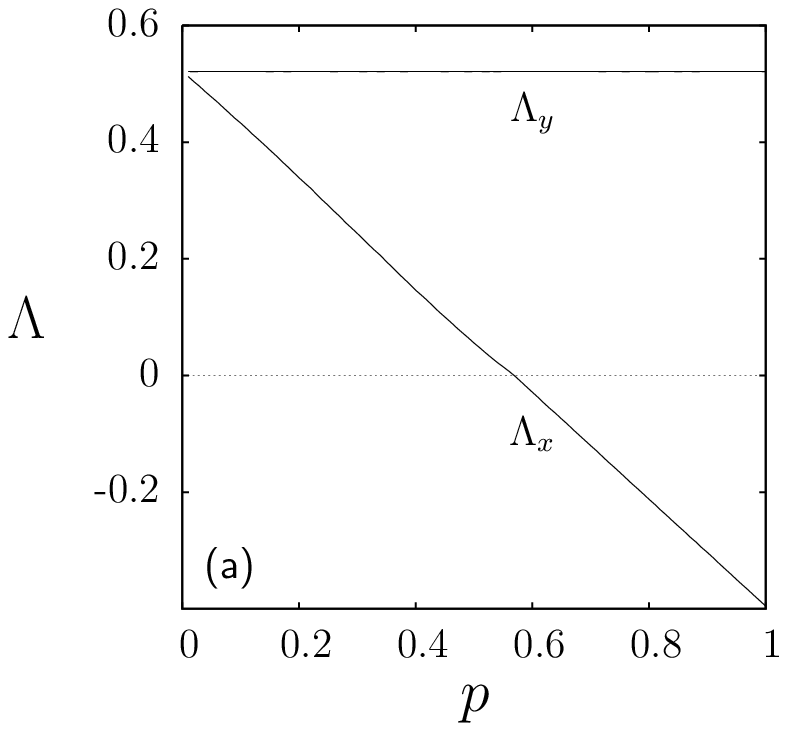} 
\includegraphics[scale=0.8]{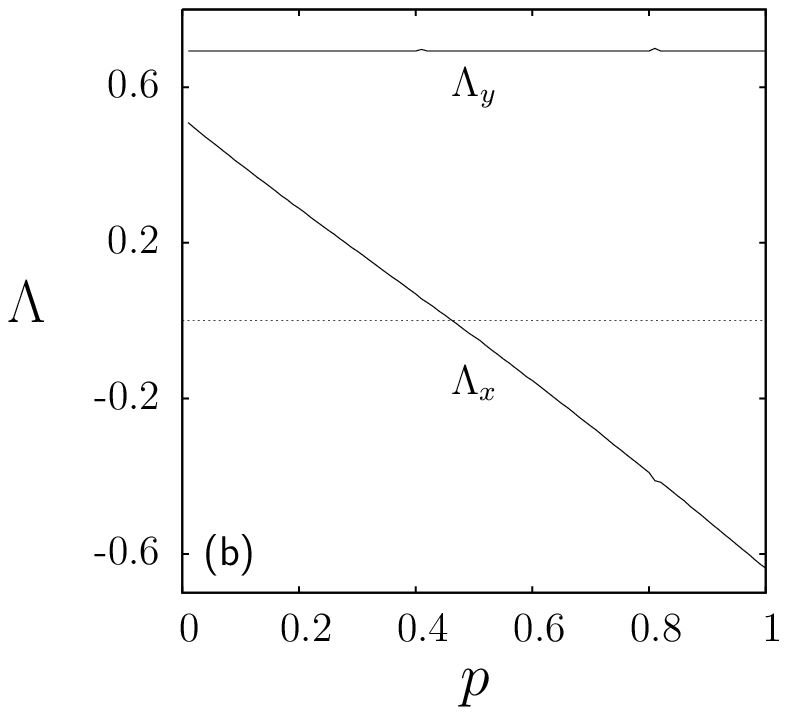}
\caption{Lyapunov exponents $\Lambda_x$ and $\Lambda_y$ as functions of $p$ for the single driven map Eq.~(\ref{2dmap}) with  $f(x)=1.2+|x|^{-0.5}$, calculated over $5 \times 10^4$ iterates after discarding $5\times 10^3$ transients for each value of $p$. (a) $g(y_t)=1.2 + |y_t|^{-0.5}$, $\epsilon=0.6$ (complete synchronization). (b)  $g(y_t)=1 - 2 y_t^2$, $\epsilon=0.8$ (generalized synchronization).}
\label{fig4}
\end{figure}

Figure~\ref{fig4} shows $\Lambda_x$ and $\Lambda_y$ as a function of $p$ for the driven map Eq.~(\ref{2dmap}) for different drives $g$.

When $g=f$, the condition $\Lambda_x<0$ implies
complete synchronization, where $x_t=y_t$. In this case we get 
\begin{equation}
{\begin{array}{l}
\Lambda_y= \lambda_f \, , \\
\Lambda_x= p \ln |1-\epsilon| + \lambda_f \, ,
 \end{array}}
\label{lyapexp}
\end{equation}
where $\lambda_f$ is the Lyapunov exponent of the map $f$.
Figure~\ref{fig2} shows the stability boundaries, given by $\Lambda_x=0$, for the completely synchronized states of the system  Eq.~(\ref{2dmap}) on the space of parameters $(p,\epsilon)$ for different orbits of a drive $g(y_t)$. 

On the other hand, if $g\neq f$, the condition $\Lambda_x<0$ corresponds to generalized synchronization, characterized by $x_t \neq y_t$. Figure~\ref{fig3} shows the stability boundaries, given by $\Lambda_x=0$, for the generalized synchronized states of the system Eq.~(\ref{2dmap}) on the space of parameters $(p,\epsilon)$ for different drives $g(y_t)$. 
In both, Figs.~\ref{fig2} and \ref{fig3}, the curves $\Lambda_x=0$ coincide with the boundaries for stability of complete and generalized synchronization, respectively, in the extended system Eq.~(\ref{int_drive}).
Thus, the condition $\Lambda_x < 0$ for generalized or complete synchronization in the single driven map Eq.~(\ref{2dmap}) is
equivalent to the condition for the stability of the synchronized state $x_t^i=x_t$, $\forall i$, in the intermittently driven system of maps Eq.~(\ref{int_drive}).

The above results are a consequence of the auxiliary system approach. Furthermore, the equivalence between 
a single driven map and a system of driven similar maps also shows that, under some circumstances, the collective behavior of an extended system of interacting elements can be inferred by considering the dynamics on a single element at the local level. As an application of this idea, we consider a partially driven system of maps defined as
\begin{equation}
{\begin{array}{l}
x^i_{t+1} = \begin{cases}
s(x^i_t,y_t), &  \, \text{with probability $p$}\\
f(x_t^i),  & \,  \text{with probability $(1-p)$}
          \end{cases} \\
y_{t+1} =  g(y_t) \, .
\end{array}}
\label{part_drive}
\end{equation}
The parameter $p$ is the probability of interaction of a map with the drive $g$ at a time $t$. 
The driven elements are randomly chosen with a probability $p$, so that not all the maps in the system receive the same external influence at all times. Thus, the average fraction of driven elements in the system Eq.~(\ref{part_drive}) at 
any given time is $p$. In comparison, the forcing of the elements in the intermittently driven system Eq.~(\ref{int_drive}) is simultaneous and uniform; each map receives the same influence from the drive $g$ at any $t$ with probability $p$.

When system Eq.~(\ref{part_drive}) gets synchronized, we have $x_t^i=x_t$. However, the  synchronized solution exists only if $g=f$. Therefore, only complete synchronization $x_t^i=x_t=y_t$ can occur in this system.

At the local level, each map in the partially driven system Eq.~(\ref{part_drive}) is subject, on the average, to an external forcing $g$ with probability $p$. Thus, the behavior of system Eq.~(\ref{part_drive}) can also be studied from the
behavior of the single, intermittently driven map Eq.~(\ref{2dmap}). In particular, if the system of maps Eq.~(\ref{part_drive})
driven with $g=f$ reaches a complete synchronized state for some values of parameters, then for this same set of parameters the single driven map Eq.~(\ref{2dmap}) subject to the same drive should eventually exhibit a synchronized state similar to that of system Eq.~(\ref{part_drive}). 
Thus, the condition $\Lambda_x < 0$ for complete synchronization in the single driven map Eq.~(\ref{2dmap}), that implies
stable synchronization in the intermittently driven system  Eq.~(\ref{int_drive}), 
is also equivalent to the condition for the stability of the complete synchronized state $x_t^i=x_t=y_t$, $\forall i$, in the partially driven system Eq.~(\ref{part_drive}). To see this, when $g=f$ we express the system Eq.~(\ref{part_drive}) 
in vector form as
\begin{equation}
 \mathbf{x}_{t+1} = \mathbf{G}_t \mathbf{f}(\mathbf{x}_t)
\label{part_drive_vect}
\end{equation}
where the $(N+1)$-dimensional vectors $\mathbf{x}_t$ and $\mathbf{f}(\mathbf{x}_t)$
have components $[\mathbf{x}_t]_i=x_t^i$ and
$[\mathbf{f}(\mathbf{x}_t)]_i=f(x_t^i)$, respectively, for $i=1,\ldots,N$, 
while $[\mathbf{x}_t]_{N+1}=y_t$ and $[\mathbf{f}(\mathbf{x}_t)]_{N+1}=g(y_t)$. 
The $(N+1)\times (N+1)$ matrix $\mathbf{G}_t$ expresses the coupling between the $N$ maps and the drive at a time $t$.
The matrix $\mathbf{G}_t$ at time $t$ possesses $pN$ randomly chosen rows, each having its components  $G_{i,j}$ $(i,j=1,2,\ldots,(N+1))$ equal to $0$, except $G_{i,i}=\epsilon$ and $G_{i,N+1}=\epsilon$.
The remaining $N-pN$ rows, and the row $(N+1)$, have their components  $G_{i,j}=0$, except $G_{i,i}=1$.
Thus, at a time $t$, the matrix $\mathbf{G}_t$  has the form
\begin{equation}
\mathbf{G}_t=
\left( \begin{array}{ccccc}
(1-\epsilon) & 0 & \cdots & \cdots & \epsilon \\
0 & 1 & 0 & \cdots & \vdots \\
\vdots & 0 & \ddots & \cdots & \vdots \\
\vdots & \vdots & 0 & (1-\epsilon) & \epsilon \\
 0 & \cdots & \cdots & 0  & 1 \\
\end{array} \right).
\label{matrix2}
\end{equation}

The long-time evolution of the vector state $\mathbf{x}_t$ is given by 
\begin{equation}
 \mathbf{x}_{t+1} = \overline{\mathbf{G}} ~ \mathbf{f}(\mathbf{x}_t).
\end{equation}
where the matrix $\overline{\mathbf{G}}$ is the geometric mean of the products of all possible configurations of $\mathbf{G}_t$
over a long time $T$,
\begin{equation}
\overline{\mathbf{G}}=\left[ \prod_{t=0}^{T-1} \mathbf{G}_t \right]^{1/T} \, .
\end{equation}
We get
\begin{equation}
\overline{\mathbf{G}}=
\left( \begin{array}{ccccc}
(1-\epsilon)^{pT} & 0 & \cdots & \cdots & K \\
0 & (1-\epsilon)^{pT} & 0 & \cdots & K \\
\vdots & 0 & \ddots & \cdots & \vdots \\
\vdots & \vdots & 0 & (1-\epsilon)^{pT} & K \\
 0 & \cdots & \cdots & 0  & 1 \\
\end{array} \right)^{1/T},
\label{aver_matrix}
\end{equation}
where $K$ is a polynomial whose terms contain products of $(1-\epsilon)$ and $\epsilon$. 

The linear stability analysis [Waller \& Kapral, 1984; Kaneko, 1990] of the complete synchronized state $f(x_t^i)=f(x_t)$ yields
\begin{equation}
\label{condition}
 \left| \alpha_k  ~ e^{\lambda_f} \right| < 1 \, ,
\end{equation}
where $\alpha_k$ $(k=0,1\ldots,N)$ are the set of eigenvalues of the matrix $\overline{\mathbf{G}}$, with $\alpha_0=1$ and $\alpha_k=(\epsilon-1)^p$ for $k>0$, having $N$-fold degeneracy. The eigenvector corresponding to $k=0$ is homogeneous. Thus perturbations of $\mathbf{x}_t$ along
this eigenvector do not destroy the coherence, and the stability condition associated with $k=0$ is irrelevant for a synchronized
state. The other eigenvectors corresponding to $k \neq 0$ are not homogeneous. Thus, condition Eq.~(\ref{condition}) with $k\neq 0$ becomes
\begin{equation}
p \ln |1-\epsilon |+ \lambda_f < 0,
\end{equation}
which is the same condition for stability of complete synchronized states in the single driven map, Eq.~(\ref{lyapexp}), when $g=f$. 
Thus, the stability boundary $\Lambda_x=0$ in Fig.~\ref{fig2} for the driven map with $g=f$ coincides with both, the boundary that separates the region where complete synchronization occurs on the space of parameters $(p,\epsilon)$ for the partially driven system Eq.~(\ref{part_drive}), and the boundary for complete synchronization in the intermittently driven system Eq.~(\ref{int_drive}).  However, generalized synchronization cannot occur in the former system.

\begin{figure}[h]
\centerline{
\includegraphics[scale=0.8]{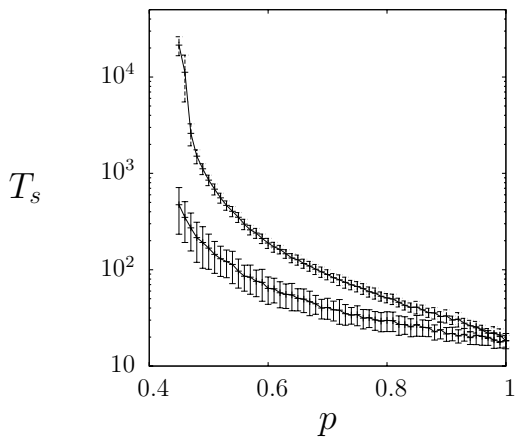}}
\caption{Average time $T_s$ to reach complete synchronization as a function of the probability $p$ for both, the intermittently driven system Eq.~(\ref{int_drive}) (bottom curve)  and the partially driven system Eq.~(\ref{part_drive}) (top curve). For both systems,  $f=1.2-|x|^{-0.5}$, $\epsilon=0.7$, $N=10^4$, and $g=f$. The error bars on each curve correspond to standard deviations resulting from $100$ realizations of random initial conditions for each value of $p$.}
\label{fig5}
\end{figure}

Although complete synchronization of chaos in both classes of driven systems, Eq.~(\ref{int_drive}) and Eq.~(\ref{part_drive}), can be characterized from the knowledge of the dynamical response of a single driven map, the transient behavior to reach such a state is different in each case.  Figure~(\ref{fig5}) shows the average time $T_s$ required to attain complete synchronization as a function of $p$ in both, a partially driven system and an intermittently driven system of maps. The times $T_s$ are larger in the first case, i.e., a spatially uniform forcing on a system is more efficient than a non-uniform one for achieving complete synchronization. 

\begin{figure}[h]
\centerline{
\includegraphics[scale=0.8]{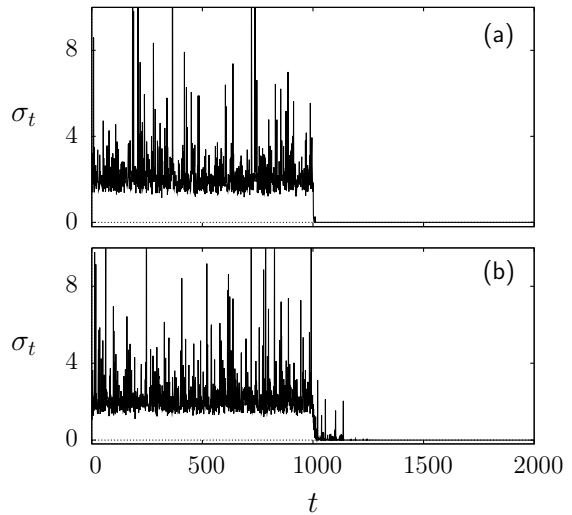}}
\caption{$\sigma_t$ vs. $t$ for (a) the intermittently driven lattice Eq.~(\ref{int_net}) and (b) the partially driven lattice
Eq.~(\ref{parcial_net}). In both cases, $f(x)=1.2+|x|^{-0.5}$, $\epsilon=0.8$, $\gamma=0.2$, $p=0.5$, $N=10^3$, and 
the drive $g=f$ is applied starting at $t=1000$. Random initial conditions are used.}
\label{fig6}
\end{figure}

Complete synchronization in networks of coupled oscillators subject to either a uniform drive $g$ with a probability $p$, or to a drive $g$ applied on a random fraction $p$ of the elements at all times, may also be equivalent.
As an illustration, consider a uniformly, intermittently driven one-dimensional lattice,
\begin{equation}
\small
\forall i, \, x^i_{t+1} = \left\lbrace 
\begin{array}{l}
 (1-\epsilon-\gamma)f(x^i_t)+ \frac{\gamma}{2}\left[ f(x_t^{i+1})+f(x_t^{i-1}) \right] +\epsilon g(y_t),
        \\    \text{with probability $p$};\\
 (1-\gamma)f(x^i_t)+ \frac{\gamma}{2}\left[ f(x_t^{i+1})+f(x_t^{i-1}) \right], \\ \text{with probability $(1-p)$}; 
\end{array}
\right. 
\label{int_net}
\end{equation}
where $\gamma$ is the local coupling parameter. The analogy 
can be established with a similar lattice subject to a drive $g(y_t)$ acting on a randomly chosen fraction $p$ of maps,
\begin{equation}
\small
x^i_{t+1} = \left\lbrace 
\begin{array}{l}
(1-\epsilon-\gamma)f(x^i_t)+ \frac{\gamma}{2}\left[ f(x_t^{i+1})+f(x_t^{i-1}) \right] + \epsilon g(y_t), \\   
 \text{with probability $p$};\\
(1-\gamma)f(x^i_t)+ \frac{\gamma}{2}\left[ f(x_t^{i+1})+f(x_t^{i-1}) \right],\\ \text{with probability $(1-p)$}. 
\end{array}
\right. 
\label{parcial_net}
\end{equation}
Periodic boundary conditions are assumed for both lattices. Figure \ref{fig6} shows
$\sigma_t$ as a function of time for both systems, Eq.~(\ref{int_net}) and Eq.~(\ref{parcial_net}), for a given example.

In summary, we have shown that the synchronization behavior of a system of chaotic maps subject to an external forcing can be inferred from the behavior of a single element in the system. The local dynamics can be seen as a single driven map; when this drive-response system reaches synchronization, the auxiliary system approach implies that an ensemble of identical maps subject to a similar drive should also synchronize. We have shown that the regions of stable complete synchronization in parameter space of two systems of identical chaotic maps subject to the same drive acting either intermittently in time, or partially in space, coincide. At the local level, the long-time dynamics of both systems can be characterized by a single chaotic map driven with a probability $p$. In addition, we have shown that complete synchronization in networks of interacting elements subject to either intermittent or partial drives can also be similar. This result suggests that these systems possess an ergodicity property.

For the intermittently driven system, the sharing of the external influence by the elements takes place only a fraction of the time. In the case of the partially driven system, 
the external drive is shared only by a fraction of elements at any time.
Thus, neither the presence of a common influence for all times, or the simultaneous sharing of the same influence by all the elements, seem essential for reaching synchronization in systems of chaotic oscillators.
What becomes crucial for achieving synchronization in both systems is
the sharing of some minimal, average information by the elements over long times.  
Future extensions of this work
should include the investigation of some quantity, such as the transfer entropy [Schreiber, 2000], for measuring this minimal amount of information required for both types of synchronization, and
the consideration of other forms of collective behaviors observed in dynamical networks, besides synchronization.

\section*{Acknowledgments}
This work was supported by Consejo de Desarrollo Cient\'{\i}fico, Human\'{\i}stico y Tecnol\'ogico, Universidad de Los Andes, Venezuela, through grant C-1579-08-05-B.

\bigskip

\noindent {\bf References} \smallskip

\noindent Abarbanel, H. D. I., Rulkov, N. F., \& Sushchik, M. M. [1996] ``Generalized synchronization of chaos: The auxiliary system approach'', \textit{Phys. Rev. E} {\bf 53}, 4528-4535.

\noindent Alvarez-Llamoza, O., Cosenza, M. G. [2008],  ``Generalized synchronization of chaos in autonomous systems'', \textit{Phys. Rev. E} \textbf{78}, 046216-1-046216-6.

\noindent Alvarez-Llamoza, O., Cosenza, M. G. \& Ponce, G. A. [2008] ``Critical behavior of the Lyapunov exponent
in type-III intermittency",  \textit{Chaos, Solitons, \& Fractals} {\bf 36}, 150-156.

\noindent Argyris, A., Syvridis, D.,  Larger, L., Annovazzi-Lodi, V.,  Colet, P., Fischer, I., 
Garcia-Ojalvo, J., Mirasso, C. R., Pesquera, L. \&  Shore, K. A. [2005] ``Chaos-based communications at high bit rates using commercial fibre-optic links'', \textit{Nature (London)} \textbf{438}, 343-346.

\noindent  Boccaletti, S., Kurths,  J., Osipov, G. ,  Valladares, D. L. \&
Zhou,  C. S. [2002] ``The synchronization of chaotic systems'', \textit{Phys. Rep.} \textbf{366}, 1-101.

\noindent Hunt,  B. R., Ott,  E. \&  Yorke, J. A. [1997] ``Differentiable generalized synchronization of chaos'', \textit{Phys. Rev. E} {\bf 55}, 4029-4034.

\noindent Kaneko, K. [1990] ``Clustering, coding, switching, hierarchical ordering, and control in networks of chaotic elements", \textit{Physica} \textbf{D 41}, 137-172.

\noindent Kapitaniak, T.,  Wojewoda, J. \& Brindley, J. [1996] ``Synchronization and desynchronization in quasi-hyperbolic chaotic systems'', \textit{Phys. Lett. A} 
{\bf 210}, 283-289.

\noindent Maritan, A., \& Banavar, J. R. [1994] ``Chaos, noise, and synchronization'', \textit{Phys. Rev. Lett.} \textbf{72}, 1451-1454.

\noindent Parlitz,  U. \&  Kokarev, L. [1999], in {\it Handbook of Chaos Control}, edited by H. Schuster, p. 271, Wiley-VCH, Weinheim.

\noindent Pecora, L. M. \&  Carroll, T. L. [1990] ``Synchronization in chaotic systems", \textit{Phys. Rev. Lett.} \textbf{64}, 821-824.

\noindent Pikovsky,  A., Rosenblum,  M.  \&  Kurths, J. [2002] \textit{Synchronization: a
universal concept in nonlinear sciences}, Cambridge University Press, Cambridge.

\noindent Rulkov,  N. F.,  Sushchik, M. M.,  Tsimring,  L. S. \& Abarbanel,  H. D. I. [1995] ``Generalized synchronization of chaos in directionally coupled systems'',  \textit{Phys. Rev. E} {\bf 51}, 980-994.

\noindent Schreiber, T. [2000] ``Measuring information transfer'', \textit{Phys. Rev. Lett.} \textbf{85}, 461-464. 

\noindent Uchida, A.,   Rogister, F.,  Garcia-Ojalvo, J. \& Roy,  R. [2005] ``Synchronization and communication with chaotic laser systems'', \textit{Prog. Opt.} \textbf{48}, 203-341.

\noindent  Waller, I. \&  Kapral, R. [1984] ``Spatial and temporal structure in systems of coupled nonlinear oscillators'', \textit{Phys. Rev. A} {\bf 30}, 2047-2055.

\noindent Zhou, B. B. \& Roy, R. [2007] ``Isochronal synchrony and bidirectional communication with delay-coupled nonlinear oscillators'', \textit{Phys. Rev. E} \textbf{75}, 026205-1- 026205-5.

\end{document}